\documentclass[a4paper,10pt,twoside]{just}

\usepackage{multicol}
\usepackage{graphicx}
\usepackage{booktabs}
\usepackage{amssymb,bm,mathrsfs,bbm,amscd}
\usepackage[tbtags]{amsmath}
\usepackage{lastpage}

\begin{document}

\fancyhead[c]{\small 10th International Workshop on $e^+e^-$ collisions from $\phi$ to $\psi$ (PhiPsi15)}
 \fancyfoot[C]{\small PhiPsi15-\thepage}

\footnotetext[0]{Received XX Oct. 2015}

\title{$L\,\to\, l\, l'\, l'\, \nu_l\, \nu_L$ in the SM and beyond\thanks{Supported by Conacyt, M\'exico}}

\author{%
      A.~Flores-Tlalpa$^{1;1)}$\email{alain@fisica.unam.mx}
\quad G.~L\'opez-Castro$^{2,3;2)}$\email{glopez@fis.cinvestav.mx} 
\quad P.~Roig$^{2;3);\dagger}$\email{proig@fis.cinvestav.mx} 
}
\maketitle

\address{%
$^1$ Instituto de F\'{\i}sica, Universidad Nacional Aut\'onoma de M\'exico, Apartado Postal 20-364, 01000 M\'exico D.F., M\'exico.\\
$^2$ Departamento de F\'isica, Centro de Investigaci\'on y de Estudios Avanzados del Instituto Polit\'ecnico Nacional, Apartado Postal 14-740, 
07000 M\'exico D.F., M\'exico.\\
$^3$ IFIC, Universitat de Val\`encia -- CSIC, Apt. Correus 22085, E-46071 Val\`encia, Spain.\\
$^\dagger$ Speaker.}

\begin{abstract}
We study the $L\to l l' l' \nu_l \nu_L$ decays ($L=\tau,\mu$; $l,l'=\mu,e$) in the Standard Model (SM) and in the effective field theory (EFT) description of the 
weak charged current at low energy, both for polarized and unpolarized $L$, keeping the $l,l'$ mass dependence. We clarify the discrepancy between 
two previous SM calculations of the decay rate improving their precision. The recent $3.5\,\sigma$ anomaly found in $\tau\to e \gamma\bar{\nu}_e\nu_\tau$ could 
be checked using our precise prediction for the $\tau\to e e e \bar{\nu}_e\nu_\tau$ decays, which shall be measured analyzing already existing data from the 
first generation B-factories. It is shown how measurements of the di-$l'$ mass distribution (with appropriate cuts) and $T$-asymmetries are able to reveal the 
corresponding lepton flavor violating (LFV) processes without neutrinos in the final state.

\end{abstract}



\begin{multicols}{2}

\section{Motivation}
(Multi)lepton lepton decays are interesting for a plethora of reasons. Historically, muon decays were essential in determining the V-A character of the weak 
charged interactions. Precise studies of deviations from this Lorentz structure motivated the development of the Bouchiat-Michel-Kinoshita-Sirlin parameters 
\cite{Michel:1949qe, Bouchiat:1957zz, Kinoshita:1957zz} which now place stringent limits on the corresponding new physics effects.\\
Radiative $L\to \ell \nu_\ell \nu_L \gamma$ decays carry information about the $\ell$ spin state which gives access to the Michel-like 
parameter $\bar{\eta}$ \cite{Pratt, Eichenberger, Fetscher}. Recent and ongoing experiments measuring the radiative muon and tau lepton decays \cite{Pocanic, 
Abdessellam} will reach up to 1$\%$ sensitivity on $\bar{\eta}$ extracted from $\tau\to \ell \nu_\ell \nu_\tau \gamma$ decays at Belle-II. In our case, where the 
photon undergoes lepton pair conversion, the corresponding Michel-like parameters \cite{Kersch1} could be measured up to 3$\%$ precision.\\
The recent measurement by BaBar of the branching ratio for the $\tau\to e \gamma \bar{\nu_e} \nu_\tau$ decays \cite{BaBar} disagrees at 3.5 $\sigma$ with the SM 
computation at NLO \cite{SMrad}, while agreement at the one sigma level is found for the muon mode (which should exhibit a larger deviation). An obvious way to test 
this anomaly shall be possible through Belle's forthcoming analysis of the $\tau\to \ell \nu_\ell \nu_\tau \gamma$ decays. An indirect way to verify it can be 
obtained by confronting our precise SM prediction of the $\tau\to 3e \bar{\nu_e} \nu_\tau$ decays \cite{OurPaper} with BaBar and Belle branching ratios (both 
first generation B-factories shall be able to measure the corresponding branching fraction $\sim4\cdot10^{-5}$).\\
Another important motivation for our study comes from the fact that the considered processes are the most important irreducible backgrounds in searches for the 
LFV $L\to l l' l'$ decays. Therefore, we aim to provide precise predictions of the SM five-lepton $L$ decays so that the TAUOLA Monte Carlo 
generator \cite{Jadach:1990mz, Jadach:1993hs} can include our matrix elements to help these searches using BaBar and Belle(-II) data.\\
Finally, let us also mention that a possible explanation to the MiniBooNe and LSND anomalies was proposed \cite{Dib} by a long-living sterile neutrino that could be 
easily searched for in the considered decays, due to its characteristic signature of a displaced vertex of several cm.\\
For all these reasons, we have performed the first SM computation of the $L\to l l' l' \nu_l \nu_L$ decays keeping daughter lepton masses (as required by the 
precise B-factories data) for the (un)polarized $L$ cases and generalize the Michel-like formalism developed in Ref.~\cite{Kersch1} accordingly, including $L$ 
polarization for the first time. We have also made the corresponding computations in the most general EFT of the weak interactions at low 
energies and discussed in detail how to overcome the backgrounds from these SM processes in searches for LFV $L\to l l' l'$ decays.

\section{SM description}

At leading order, the considered decays proceed through photon emission (followed by lepton pair conversion) from the charged leptons in the well-known 
$L^-\to \ell^- \bar{\nu}_\ell \nu_L$ processes. Our notation for momenta is
\begin{equation}
L^-(Q) \to \ell^-(p_1)\;\ell'^-(p_2)\;\ell'^+(p_3)\;\bar{\nu}_{\ell}(p_4)\;\nu_{L}(p_5)\ .
\end{equation}
The masses of the charged leptons will be denoted by $M,\ m_1,\ m$ ($m_2=m_3=m$), where $M$ stands for the mass of the decaying lepton. The two diagrams in 
the $l\neq l'$ case double in the identical particle case giving rise to two extra terms that are obtained from the original ones by (antisymmetric) exchange of 
$p_1$ and $p_2$.\\
Since neutrinos are massless, the decay matrix element can be written as (both for the unpolarized and polarized $L$ cases)
\begin{equation}
\overline{|{\cal M}|^2}=\frac{1}{2}\sum_{\rm pols} |{\cal M}|^2 = {\cal T}_{\alpha\beta}p_4^{\alpha}p_5^{\beta}\ ,
\label{p4p5}
\end{equation}
where ${\cal T}_{\alpha\beta}^s$ will denote the polarized case, with $s$ referring to the $L$ helicity.\\
Being five-body decays, the corresponding squared (unpolarized) amplitude $\overline{|{\cal M}|^2}$ depends upon 8 independent kinematical variables that we choose 
following the invariants proposed in Ref.~\cite{Kumar}. Since precise numerical integrations in the multi-particles case are far from trivial, instead of 
integrating over phase space as in Ref.~\cite{Dicus-Vega} (employing the factorized expression eq.~(\ref{p4p5}) to integrate neutrinos phase space first and then 
integrating over the remaining variables), our integration following the optimized variables in Ref.~\cite{Kumar} allows for a completely independent evaluation 
of the observables.\\
Analytic results for the matrix elements in the (un)polarized case can be found in the appendices of our paper \cite{OurPaper}.\\
Branching ratios for the different modes are shown in Table \ref{Table1}, where our predictions are compared to previous results and available experimental data. 
The quoted errors arise from VEGAS \cite{Lepage:1977sw} integration, used in our numerical calculations and in Ref.~\cite{Dicus-Vega}. This comparison shows good 
agreement between our results and those in Ref.~\cite{Dicus-Vega}, while sizable discrepancies are found with the results in Ref.~\cite{CLEO}. Moreover, the closer 
agreement (at the one $\sigma$ level) with Ref.~\cite{Dicus-Vega} in the modes with two or three electrons than in the modes with two or three muons (about three 
$\sigma$) could be attributed to the different tau mass values used in both works (the heavier the daughter particles, the larger the effect). Remaining tiny 
differences can be explained because of the overall effect of different tau lifetimes employed in these analyses.

\end{multicols}
\begin{table*}[h!]
\begin{center}
\begin{tabular}{|c||c|c|c|c|}
\hline 
Channel & Ref. \cite{Dicus-Vega} & Ref. \cite{CLEO} & This work & PDG \cite{pdg2014}\\
\hline\hline
BR($\tau^- \to e^-e^+e^- \bar{\nu}_e\nu_{\tau}$)$\times 10^5$ & $4.15\pm0.06$ & $4.457\pm0.006$ & $4.21\pm0.01$ & $2.8\pm1.5$\\
BR($\tau^- \to e^- \mu^+\mu^- \bar{\nu}_e\nu_{\tau}$)$\times 10^7$ & $1.257\pm0.003$ & $1.347\pm0.002$ & $1.247\pm0.001$ & -\\
BR($\tau^- \to \mu^-e^+e^-\bar{\nu}_{\mu}\nu_{\tau}$) $\times 10^5$ & $1.97\pm0.02$ & $2.089\pm0.003$ & $1.984\pm0.004$ & $<3.6$\\
BR($\tau^- \to \mu^-\mu^+\mu^- \bar{\nu}_{\mu}\nu_{\tau}$) $\times 10^7$ & $1.190\pm0.002$ & $1.276\pm0.004$ & $1.183\pm0.001$ & -\\
BR($\mu^- \to e^-e^+e^-\bar{\nu}_e\nu_{\mu}$) $\times 10^5$ & $3.60\pm0.02$ & $3.605\pm0.005$ & $3.597\pm0.002$ & $3.4\pm0.4$\\
\hline 
\end{tabular} 
\caption{\label{Table1}\small{Branching ratios for the five-lepton decays of taus and muons. Some of the previous calculations are shown 
in the middle columns. Experimental data are rather scarce.}}
\end{center}
\end{table*}
\begin{multicols}{2}

The normalized differential decay width (with respect to the decaying lepton decay width) versus $q^2=m_{\ell'^+\ell'^-}^2$ 
is plotted in fig.~\ref{fig1} for the considered channels. All of them peak at low $m_{\ell'^+\ell'^-}^2$ values given the low virtuality of the photon close to 
threshold.
\vspace{4mm}
\begin{center}
\includegraphics[width=5cm,angle=-90]{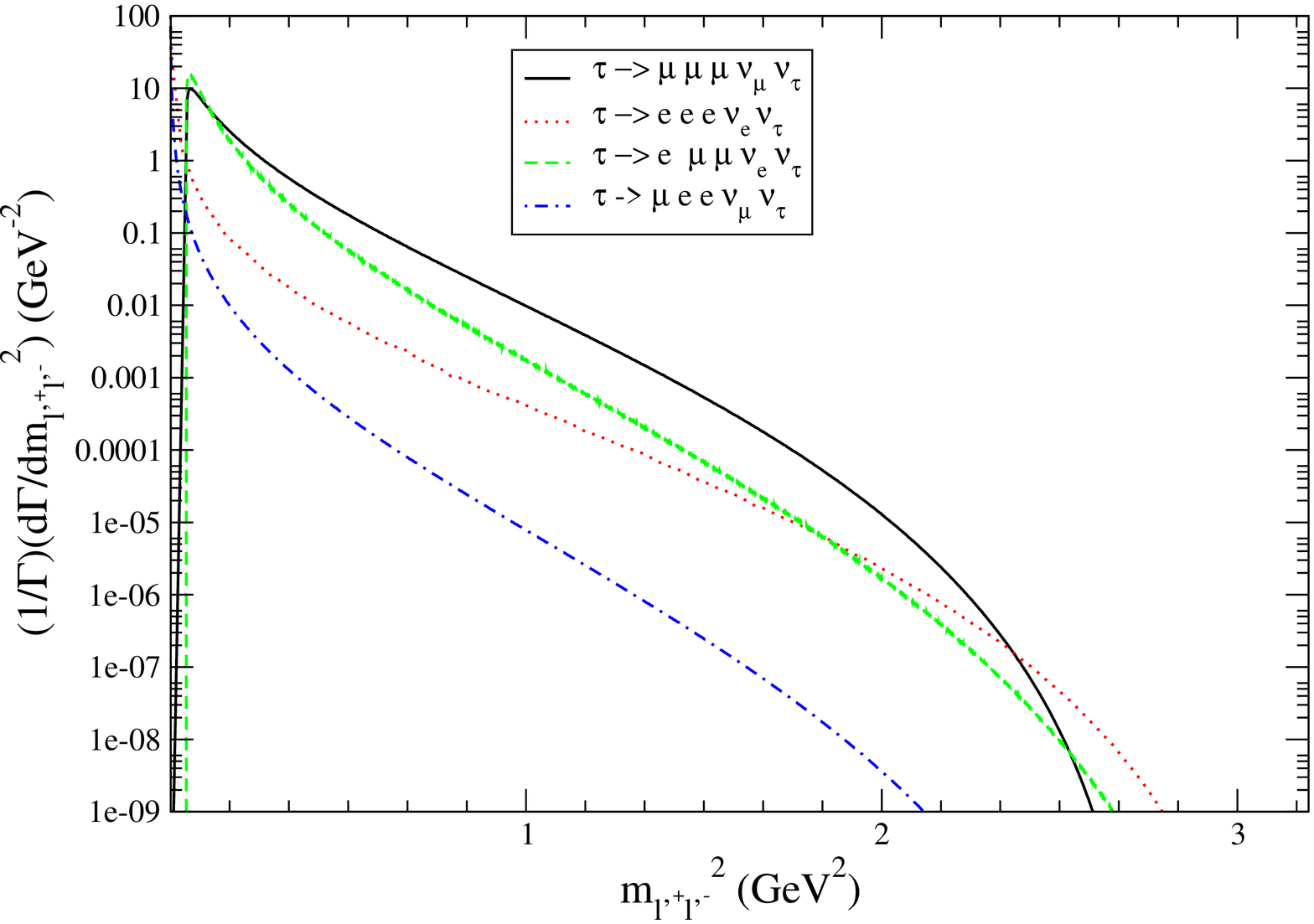}
\figcaption{\label{fig1} \small{Differential decay width (normalized to the partial decay width) versus $q^2=m_{\ell'^+\ell'^-}^2$.}}
\end{center}

Inclusion of $L$ spin effects is straightforward. After integrating over neutrinos phase space, the differential decay width reads

\begin{equation}\label{polarizedcase}
{\cal T}_{\alpha\beta}^{s} I^{\alpha\beta}(P) = e^4 G_F^2 \Big[ F \, - \,L\,\vec{{\bf p}}_1\cdot \vec{\bf s} \, -\, G_1\,\vec{\bf p}_2\cdot \vec{\bf s}\,-\, G_2\,\vec{\bf p}_3\cdot \vec{\bf s} \Big] \ ,
\end{equation}
where the generalized form factors $F$, $L$, $G_1$ and $G_2$ depend only upon scalar products of charged leptons momenta.

The most convenient writing of the differential decay width is in terms of the normalized charged daughter lepton energies and solid angles:
\begin{equation}\label{polarizedcase2}
\frac{d\Gamma_5}{dx_1 d\Omega_1 dx_2 d\Omega_2 dx_3 d\Omega_3} = \frac{M^2 |\vec{{\bf p}}_1| |\vec{{\bf p}}_2| |\vec{{\bf p}}_3|}{3\cdot 2^{21} \pi^{10}}{\cal T}_{\alpha\beta}^{s} I^{\alpha\beta}(P)\,,
\end{equation}
with $x_i=2E_i/M$ ($i=1,2,3$). In our paper \cite{OurPaper}, we have derived the first formulation in terms of Michel-like parameters for these decays in the 
polarized $L$ case keeping charged leptons masses throughout (which are first kept in our work in the unpolarized case).

\section{EFT analysis}

The precise predictions given in table \ref{Table1} allow for a stringent test of the $V-A$ character of the charged weak interactions. With this purpose, we have 
made an EFT analysis (valid at energy scales much smaller than that of electroweak symmetry breaking, as it is the case in $L$ decays) of these 
processes.\\
The most general lepton-flavor conserving Lagrangian describing four-lepton interactions (local and without derivatives) is \cite{Michel:1949qe, Bouchiat:1957zz, 
Fetscher, Kersch1, Kinoshita:1957zz, Scheck:1977yg, Pich:1995vj}
\begin{equation}\label{Lag_gral}
 \mathcal{L}\,=\,-\frac{4G_F}{\sqrt{2}}\sum_{i,\lambda,\rho} g^i_{\lambda\rho} \left[\overline{\ell'_\lambda} \Gamma^i (\nu_{\ell'})_\xi\right]\left[\overline{(\nu_\ell)_\kappa} \Gamma_i \ell_\rho\right]\,.
\end{equation}
The Lorentz structure of the weak currents is indicated by means of $i=S,\,V,\,T$; $\Gamma^S=I$, $\Gamma^V=\gamma^\mu$, $\Gamma^T=\sigma^{\mu\nu}/\sqrt{2}$ and 
$\lambda,\rho=L,R$ stand for the charged lepton's helicity (analogously $\xi$ and $\kappa$ for neutrinos). The 10 independent complex coefficients (4 scalar, 4 
vector and the two tensor couplings $g^T_{LR}$ and $g^T_{RL}$) give rise to 19 independent real couplings (removing an unphysical global phase).\\
The relevant sum over the $|g^i_{\lambda\rho}|^2$ in eq.~(\ref{Lag_gral}) is fixed by the total decay width \cite{Fetscher}
\begin{equation}\label{eq_normalization_gral}
 1\,=\,\frac{1}{4}\sum_{\lambda,\rho} |g^S_{\lambda\rho}|^2+\sum_{\lambda,\rho} |g^V_{\lambda\rho}|^2 + 3 \big( |g^T_{RL}|^2 + |g^T_{LR}|^2 \big)\,.
\end{equation}
In the Standard Model, $|g^V_{LL}|=1$ is the only non-vanishing coupling. Moreover, in the limit of massless left-handed neutrinos, only $g^S_{RR}$ survives, 
leaving the interference between these two contributions as the most promising signature of new physics affecting these decays.\\
The presence of the tau polarization vector could \textit{a priori} allow for six spin-dependent generalized form factors in such a way that the 
eq.~(\ref{polarizedcase}) would generalize to
\begin{eqnarray}\label{polarized-gral-case}
\mathcal{T}_{\alpha\beta}^{s} I^{\alpha\beta}(P)&=& e^4 |G_F|^2 \Big[ F \, - \, L\,\vec{\bf p}_1\cdot \vec{\bf s} \, -\, G_1\,\vec{\bf p}_2\cdot \vec{\bf s}\,-\, G_2\,\vec{\bf p}_3\cdot \vec{\bf s}  \\   
&& +H_1\vec{\bf s}\cdot (\vec{\bf p}_1\times \vec{\bf p}_2) 
 + H_2\vec{\bf s}\cdot (\vec{\bf p}_2\times \vec{\bf p}_3) + H_3\vec{\bf s}\cdot (\vec{\bf p}_1\times \vec{\bf p}_3) \Big] \ .\nonumber
\end{eqnarray}
However, since CP violation (through T violation, in this case) effects are suppressed (in the case of Dirac neutrinos) by a lepton analog of the Jarlskog \cite{Jarlskog} 
invariant involving, in particular, the product $(m_{\nu_1}^2-m_{\nu_2}^2)(m_{\nu_1}^2-m_{\nu_3}^2)(m_{\nu_2}^2-m_{\nu_3}^2)$; again neutrino masses make this 
effect unmeasurable and the only non-vanishing spin-dependent form factors in this limit are $L$, $G_1$ and $G_2$. Their corresponding expressions can be found 
in the appendices of our paper \cite{OurPaper}.\\
On the contrary, since CP violation effects are not suppressed in the LFV $L\to \ell' \ell^+\ell^-$ decays \cite{Okada:1999zk, Kitano:2000fg} \footnote{A T 
asymmetry as large as $15\%$ can be found \cite{Okada:1999zk} for $\mu\to 3e$ decays in specific GUTs.}, dedicated searches 
of T violating effects in processes with a lepton decaying into three charged leptons could unveil LFV.

\section{LFV $L\to l l' l'$ decays}

The considered five-lepton lepton decays constitute the most difficult irreducible background in the searches for LFV $L\to l l' l'$ processes, particularly in 
the configuration where both neutrinos carry away so little energy that they cannot be detected. It is thus interesting to compare our predictions for different 
observables to the expected signal of the LFV processes in order to gain insight for the experimental searches of the latter.\\
In this spirit, we have considered the three benchmark scenarios proposed in Ref. \cite{Celis:2014asa} for dipole, scalar and vector interactions originating the 
LFV processes. We have considered the differential decay width as a function of one of the lepton energies and of the same- and opposite-sign lepton pair. It is 
concluded that the distributions in $m_{\ell'\ell'}^2$ are best suited to split signal from background with a moderate cut between $0.5$ 
and $1$ GeV$^2$ depending on the considered tau decay channel \cite{OurPaper}. In the case of muon decays, huge statistics and an exquisite control of SM 
backgrounds will be needed to compensate the tiny signal to background ratio given in table \ref{Table2}.

\vspace{4mm}
\begin{center}
\includegraphics[width=5cm,angle=-90]{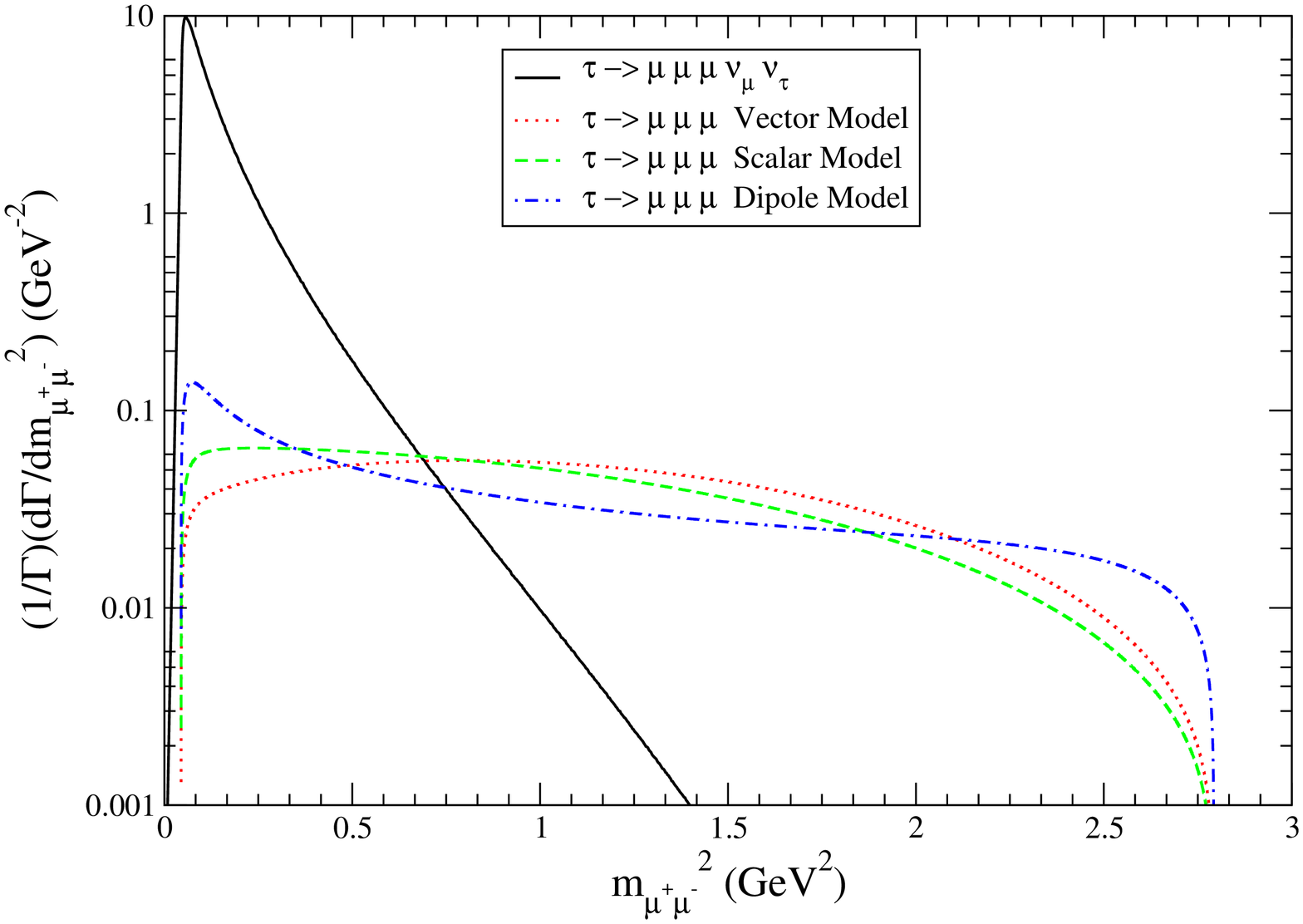}
\figcaption{\label{fig2} \small{Comparison of opposite-sign lepton-pair invariant mass distribution in five-body and LFV decays of tau leptons. Three benchmark 
scenarios of new physics in $\tau\to3\mu$ are used, according to Ref. \cite{Celis:2014asa}, and $S/B\sim0.1$ (about the current upper limit) is assumed.}}
\end{center}

\end{multicols}
\begin{table*}[h!]
\begin{center}
\begin{tabular}{|c||c|c|c|}
\hline 
Channel & Current upper limit (UL) \cite{pdg2014, Amhis:2014hma} & S/B (UL) & Expected UL \cite{Hayasaka:2010np}\\
\hline\hline
BR($\tau^- \to e^-e^+e^-$) & $1.4\cdot10^{-8}$ & $\sim3\cdot10^{-4}$ & $\sim10^{-9}$\\
BR($\tau^- \to e^- \mu^+\mu^-$) & $1.6\cdot10^{-8}$ & $\sim0.1$ & $\sim10^{-9}$\\
BR($\tau^- \to \mu^-e^+e^-$) & $1.1\cdot10^{-8}$ & $\sim6\cdot10^{-4}$ & $\sim10^{-9}$\\
BR($\tau^- \to \mu^-\mu^+\mu^-$) & $1.2\cdot10^{-8}$ & $\sim0.1$ & $\sim10^{-9}$\\
BR($\mu^- \to e^-e^+e^-$) & $1.0\cdot10^{-12}$ & $\sim3\cdot10^{-8}$ & $\sim10^{-16}$\\
\hline 
\end{tabular} 
\caption{\label{Table2}\small{Current and expected UL and signal to background ratios in LFV $L^-\to \ell^-\ell'^+\ell'^-$ searches.}}
\end{center}
\end{table*}
\begin{multicols}{2}

\section{Conclusions and outlook}
We have made the first analysis of the five-lepton lepton decays keeping daughter masses for the (un)polarized parent lepton cases. Our results agree with one 
of the earlier references in the literature improving substantially the precision. We have derived analytic expressions for the Michel-like parameters useful for 
accurate tests of the V-A character of the charged weak interactions including decaying lepton spin effects. Our precise results would also allow to test the 3.5 
$\sigma$ anomaly found by BaBar in $\tau\to e\gamma\bar{\nu_e}\nu_\tau$ decays by analyzing (with first generation B-factories data) the $\tau\to 3e\bar{\nu_e}\nu_\tau$ 
processes. We have also extended our computations to an EFT description of these at low energies and show that, even in this case, 
$T$-violation is suppressed by the tiny neutrino masses. Therefore, a wise way of searching for LFV $L\to\ell\ell'\ell'$ is to measure 
non-vanishing T-odd correlations involving charged leptons, which would point to the beyond the SM process immediately. Finally, we have shown that 
reasonable cuts in $m_{\ell'\ell'}$ are very efficient in splitting signal from background in these new physics searches.\\

\acknowledgments{We thank the Organizing Committees and Secretariat of PhiPsi15 for the very interesting and fruitful Hefei conference. P.~R. acknowledges 
financial support from Dpto. de F\'isica del Cinvestav for attending this workshop. This work was partly funded by Conacyt (M\'exico) and A.F.T. was supported 
in part by DGAPA-UNAM. The work of G.L.C. has been supported in part by EPLANET project and by the Spanish Government and ERDF funds from the EU Commission 
[Grants No. FPA2011-23778, FPA2014-53631-C2-1-P, SEV-2014-0398] and by Generalitat Valenciana under Grant No. PROMETEOII/2013/007. 
We appreciate very much discussions with Denis Epifanov and Mart\'{\i}n Gonz\'alez-Alonso concerning this research.}

\end{multicols}

\vspace{10mm}

\vspace{-1mm}
\centerline{\rule{80mm}{0.1pt}}
\vspace{2mm}

\begin{multicols}{2}

\end{multicols}

\clearpage

\end{document}